\documentclass[useAMS,usenatbib]{mn2e}
\usepackage{times}
\usepackage{graphicx}
\usepackage{amsmath}
\usepackage{natbib}
\bibpunct{(}{)}{;}{a}{}{,}
\usepackage{psfig}
\title[Globular clusters in Fornax Cluster galaxies]{Globular cluster 
systems in low-luminosity early-type galaxies near the Fornax Cluster centre}

\author[L. P. Bassino, T. Richtler and B. Dirsch]{Lilia P. Bassino$^{1}$\thanks{
E-mails:\,lbassino@fcaglp.unlp.edu.ar\,(LPB); tom@mobydick.cfm.udec.cl (TR); 
bdirsch@cepheid.cfm.udec.cl (BD)}, 
Tom Richtler$^{2\star}$ and Boris Dirsch$^{2\star}$\\ 
$^{1}$Facultad de Ciencias Astron\'omicas y Geof\'{\i}sicas,
       Universidad Nacional de La Plata,
       Paseo del Bosque S/N, 1900-La Plata,\\
       Argentina; and IALP-CONICET\\
$^{2}$Universidad de Concepci\'on, Departamento de F\'{\i}sica,
      Casilla 160-C, Concepci\'on, Chile}

\begin{document}

\date{Accepted . Received ; in original form }

\pagerange{\pageref{firstpage}--\pageref{lastpage}} \pubyear{2005}

\maketitle                              

\label{firstpage}

\begin{abstract}
We present a photometric study of the globular cluster systems of the Fornax 
cluster galaxies NGC\,1374, NGC\,1379, and NGC\,1387. The data consists of 
images from the wide-field MOSAIC~~Imager of the CTIO 4-m 
telescope, obtained with Washington $C$ and Kron--Cousins $R$ filters. 
The images cover a field of 36 $\times$ 36 arcmin, corresponding to 
200 $\times$ 200\,kpc at the Fornax distance. Two of the galaxies, NGC\,1374 and 
NGC\,1379, are low-luminosity ellipticals while NGC\,1387 is a  
low-luminosity lenticular.
Their cluster systems are still embedded in the cluster system of NGC\,1399. 
Therefore the use of a large field is crucial and some differences to previous 
work can be explained by this.  The colour distributions of
all globular cluster systems are bimodal.   
NGC\,1387 presents a particularly distinct separation between red and blue clusters
and a overproportionally large population of red clusters. The radial distribution 
is different for blue and red clusters, red clusters being  more concentrated 
towards the respective galaxies. The different colour and radial distributions 
point to the existence of two globular cluster subpopulations in these galaxies. 
Specific frequencies are in the range $S_{N}= 1.4-2.4$, smaller than the typical 
values for elliptical galaxies. These galaxies might have 
suffered tidal stripping of blue globular clusters by NGC\,1399. 
\end{abstract}

\begin{keywords}
galaxies: individual: NGC 1374, NGC 1379, NGC 1387  
-- galaxies: clusters: general -- galaxies: elliptical and lenticular, cD  
-- galaxies: star clusters -- galaxies: photometry -- galaxies: haloes
\end{keywords}
     
%=====================================================================
\section{Introduction}

There is an extensive literature on  globular cluster systems 
(GCSs) of large galaxies located in the centre of clusters (e.g. the 
reviews from \citealt{ash98} or \citealt{har01}) but low-luminosity 
galaxies with their own GCSs are also present in such environments. 
The centres of galaxy clusters are interesting regions for studying  
possible interactions between the central massive galaxy 
and its less massive galactic neighbours. These interactions 
may also influence the properties of their cluster systems.  

The nearby Fornax cluster offers the best opportunity for studying such a 
scenario due to the huge GCS that surrounds its central giant galaxy, 
NGC\,1399 \citep[~and references therein]{dir03a}, and the presence of 
several low-luminosity galaxies in its neighbourhood.
The GCS of NGC\,1404, one of these less massive galaxies, has been 
studied by \citet{ric92} and, more recently, by \citet{for98} 
and \citet{gri99}. The latter authors identified two GC 
subpopulations: metal-poor GCs (blues) and metal-rich ones (reds), 
the metal-rich clusters being more centrally  concentrated than the 
metal-poor ones. The low specific frequency $S_{N}$ 
\citep[as defined by][]{har81} of this system has been understood  as  
a deficiency of GCs, probably stripped in an interaction process with  
NGC\,1399 (Forbes, Brodie \& Grillmair 1997). Numerical simulations 
were performed by  \citet{bek03} to analyse this scenario of tidal 
stripping and accretion.

One may speculate if GCs lost by low-luminosity galaxies are really 
captured by the central galaxy or just remain unbound within the galaxy 
cluster, like the proposed intracluster GCs \citep{whi87,wes95,bas03}.    

Our aim with the present paper is to study more early-type galaxies 
in the Fornax cluster: NGC\,1387 (type S0), NGC\,1379 (E0) and NGC\,1374 
(E1). These galaxies are located at angular distances from NGC\,1399 of 19, 
29 and 41 arcmin, respectively. 

The GCS of NGC\,1387 has first been studied with  CCD techniques by \citet{gre90} 
who obtained a surprisingly red mean colour $(B-V) = 1.0$. Later, \citet{koh96} and 
\citet{kis97a} performed $V$ and $I$ photometry of its GCS as well as  
of  NGC\,1374 and NGC\,1379. The former paper focused on the GC luminosity 
functions, the latter on the colour, spatial distributions, and total numbers  
of GCs. They concluded that the galaxies have 
between 300 and 500 GCs ($S_{N} \approx 4 \pm 1$) and that their $(V-I)$ 
colours are redder than those of the Milky Way GCs. However, they found no 
evidence for multiple populations, but simply pointed to a wide range in 
metallicities. 
None of the GCSs seem to be elongated and the density profiles follow the 
respective galaxy light profiles. The  background was  
determined locally within the frames of 7.4 $\times$ 7.4 arcmin used 
in these papers. 

The GCS of NGC\,1379 has been the target of a HST-study \citep{els98} 
combining HST data (B band) with  ground-based observations in 
the $B$ and $I$ bands. They estimated the total population in NGC\,1379 
to be 440 GCs, 
based on a Gaussian luminosity function with mean $<B>$ = 24.95 and 
$\sigma_{B}$ = 1.55 mag. Again, no evidence for bimodality was  found in the $(B-I)$ 
colour distribution. A specific frequency $S_{N}= 3.5$ was obtained up to 
a radius of 70 arcsec . The background corrections  
were measured in a field located 1\fdg4 from the Fornax cluster centre 
\citep[see also][]{gri99}.  

The apparent lack of multiple GC subpopulations in these 
three galaxies was taken by \citet{kis98} as an 
argument to set several constraints on the process of formation of elliptical 
galaxies. The impact posed by the existence of GCs subpopulations on the 
studies of origin and evolution of galaxies is one of the reasons why 
we think that their GCSs deserve to be re-examined with a better metallicity 
resolution. 
    
On the basis of data provided by  \citet{kis97a}, \citet*{for97} suggest that 
the early-type galaxies in Fornax, located close to NGC\,1399, might have been 
tidally stripped off their GCs. They note that their $S_{N}$ values 
increase with projected distance from the cluster centre. Moreover, their GC surface 
density slopes are similar to that of the underlying starlight, perhaps as 
a result of losing the outermost clusters. Both effects have also been studied 
through dynamical models by \citet{bek03}. 

On the other hand, \citet{cap01} use the \citet{els98} data for NGC\,1379 and 
the \citet{for98} data for NGC\,1399 and NGC\,1404 (all from HST) to show 
that the surface brightness profiles of the underlying galaxies are steeper 
than the GC profiles. They 
propose that GCs from the inner regions might have been lost as a consequence 
of dynamical friction and tidal interaction with the galactic nuclei.
  
We re-investigate the GCSs in NGC\,1387, NGC\,1379 and NGC\,1374, 
on the basis of new MOSAIC images in the Washington photometric system.  
This system provides a metallicity resolution sufficiently good for a firm 
detection of GCs subpopulations. The large field improves the 
determination of the background (partly due 
to the NGC\,1399 GCs) that may otherwise alter the radial density profiles 
and luminosity functions of these low number GCSs.

The analysis of the GCSs of our three target galaxies  
is organized as follows: Section 2 describes the observations 
and the adopted criteria for the globular cluster candidates' selection. 
In Section 3 we analyse the colour and spatial distributions, the luminosity 
functions of the GCSs and also estimate the total populations. A discussion 
of the results is provided in Section 4. Finally, a summary of the results 
and their implications are provided in Section 5.

%=======================================================================
\section{Observations and Data Reduction}

\subsection{Observations and point sources selection}

The observations were performed with the MOSAIC camera (8 CCDs mosaic imager) 
mounted at the prime focus of the 4-m Blanco telescope at the Cerro Tololo 
Inter--American Observatory (CTIO), during 2001 November 17--19. 
One pixel of the MOSAIC wide-field camera subtends 0.27 arcsec on the sky, 
which results in a field of 36 $\times$ 36 arcmin (200 $\times$ 200 
kpc at the Fornax distance). All three target galaxies were covered  
by one MOSAIC field. For more information on the MOSAIC camera we refer to 
its homepage 
({\it http://www.noao.edu/kpno/mosaic/mosaic.html}).

Kron-Cousins $R$ and Washington $C$ filters were used. We selected the 
$R$ filter instead of the original Washington $T_1$  as \citet{gei96a}  
has shown that the Kron-Cousins $R$ filter is more efficient than $T_1$, 
and that $R$ and $T_1$ magnitudes are very similar, with just a very small 
colour term and zero-point difference ($R - T_1 \approx -0.02$). 

In order to fill in the gaps between the 8 individual MOSAIC chips, the data 
were dithered taking three images in $R$ with exposure times of 600 s each, 
and six images in $C$ with exposures of 1\,200 s each. Fig.~1 shows a 
$C + R $ combined image of the MOSAIC field where the three galaxies have 
been identified. Their basic data are given in Table~1.

\begin{figure}
\resizebox{\hsize}{!}{\includegraphics{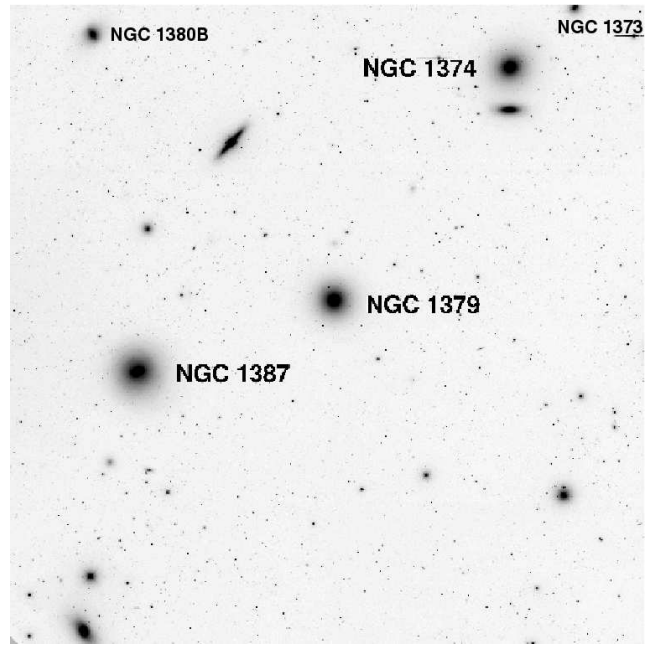}}
\caption{$C + R $ combined image of the MOSAIC field. Labels indicate 
the three target galaxies. North is up and East to the left.}
\end{figure}

\begin{table*} 
\parbox{6.0cm}{\vskip0.4cm
\caption{General data for the target galaxies in the Fornax 
cluster. Total blue apparent magnitudes ($B_{\rmn T}$) and heliocentric radial 
velocities (RV) were taken from the RC3 catalog, and reddening values 
from \citet*{sch98}.}}
\parbox[b]{11.6cm}{
\flushright
\begin{tabular}{ccccccc}
\hline
Galaxy & RA(J2000) & DEC(J2000) & Type & $B_{\rmn T}$ &  RV\,[km\,sec$^{-1}$] & $E(B-V)$\\
\hline
NGC 1374 & $3^h 35^m 17^s$&$-35\degr 13\arcmin 35\arcsec$&E1& 12.00 &$1351\pm15$&0.014\\
NGC 1379 & $3^h 36^m 03^s$&$-35\degr 26\arcmin 28\arcsec$&E0& 11.80 &$1376\pm19$&0.012\\
NGC 1387 & $3^h 36^m 57^s$&$-35\degr 30\arcmin 24\arcsec$&S0& 11.68 &$1328\pm24$&0.013\\
\hline
\end{tabular}}
\end{table*}

The MOSAIC data were reduced using the {\it mscred} package within 
IRAF\footnote{IRAF is distributed by the National Optical Astronomy 
Observatories, which is operated by AURA, Inc.\ under contract to the 
National Science Foundation.}. The software corrects for the variable 
pixel scale across the CCD which might cause a 4 per cent variability 
in point sources brightness (from the centre to the corners).
The final $R$ image resulted with remaining sensitivity variations up 
to 0.8 per cent, and the final $C$ image up to 1.5 per cent. The seeing in 
these final images was 1 arcsec on the $R$ frame, and 1.2 arcsec on 
the $C$ frame.

The halo light of each galaxy 
was subtracted by means of a ring median filter with an inner boundary of  
1.3 arcsec and an outer boundary of 10.7 arcsec. In previous work based
on MOSAIC images \citep{dir03a,dir03b}, we have already checked
that this process does not affect the photometry. Anyway, we also have  
compared the magnitudes obtained with and without filtering for several 
point sources on different positions on the image. The differences were 
negligible as compared to the photometric errors  (see
below).

The photometry was done with DAOPHOT within IRAF. For the first search we
used DAOFIND on a combined $C + R$ image and  20\,856
unresolved and extended objects were detected in the whole field.  
In the final $C$ and $R$ images, a second order variable PSF was 
derived using about 100 evenly  distributed stars per frame, which was 
adjusted to the sources through the ALLSTAR task. The estimated 
errors in the aperture corrections between the PSF radius and a  
15\,pix radius were $\approx$ 0.02 and 0.01 mag for the $C$ and $R$ images,
respectively.

Several tests were carried out before deciding how to select the point sources. 
We compared the results obtained using the stellarity index estimated by the 
SExtractor software \citep{ber96} which is a reliable tool for separating 
point sources and extended objects \citep[see, for instance,][]{non99},  
adopting an stellarity index between 0.4 and 1 for point sources 
\citep[following][]{dir03a}, with the results obtained from DAOPHOT 
using different options for the parameters $\chi$ and sharpness of 
the ALLSTAR task. 
We finally adopted as point sources those objects selected with the 
ALLSTAR parameters $\chi < 2$ and sharpness between -0.5 and 0.5 
\citep[see, for instance,][]{rei96}. 

In this way, out of the total number of objects originally detected 
in the field, 6\,520 point sources were selected, populating a magnitude 
range of $18.5 \la T_1 \la 24.5$.

%--------------------------------------------------------------------
\subsection{Photometric calibration and globular clusters selection}

The photometric calibration of our observations already has been performed
by \citet{dir03a} so we conveniently can use their relations. 
 Fields of standard 
stars were selected from the list of \citet{gei96a}. Four or 
five fields, with roughly 10 standard stars located on each of them, 
were obtained in each night. Care was taken to cover 
a large range in airmasses (1.0 to 1.9 approximately). 

In order to calculate the standard magnitudes from the instrumental ones, 
we have used the equations already obtained by \citet{dir03a} for this run, 
which we reproduce here:

\begin{eqnarray*}
\mathrm{T_1} & = & \mathrm{R}_\mathrm{inst}+(0.72\pm0.01)-(0.08\pm0.01)X_\mathrm{R}\\
        & & \mbox{} +(0.021\pm0.004)(\mathrm{C}-\mathrm{T_1})\\
\mathrm{C} & = &  \mathrm{C}_\mathrm{inst}+(0.06\pm0.02)-(0.30\pm0.01)X_\mathrm{C}\\
        & & \mbox{} +(0.074\pm0.004)(\mathrm{C}-\mathrm{T_1})
\end{eqnarray*}

These equations also include the zero-point difference to transform $R$ into 
$T_1$ magnitudes (as explained in Section 2.1) so, in the rest of this paper, 
we will use $T_1$ magnitudes, not $R$.  
The photometric errors for the point sources are shown in  Fig.~2.

The colour-magnitude diagram for all point sources in the field is
plotted in Fig.~3. Globular cluster candidates can be clearly 
distinguished in the range $0.8 \leq (C-T_1) \leq 2.3$ mag. Objects bluer 
than $(C-T_1) = 0.8$ and fainter than  $ T_1 = 23 $ are mostly compact 
background galaxies while objects redder than $(C-T_1) \approx 2.8 $  
are mainly foreground stars.

\begin{figure}
\resizebox{\hsize}{!}{\includegraphics{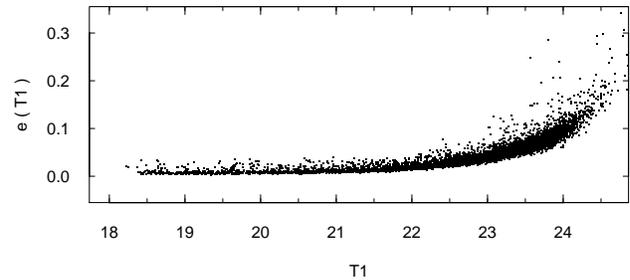}}
\caption{Photometric errors in $ T_1$ from DaoPhot plotted against the 
$ T_1$ magnitude for all point sources in the MOSAIC field.
}
\end{figure}

\begin{figure}
\resizebox{\hsize}{!}{\includegraphics{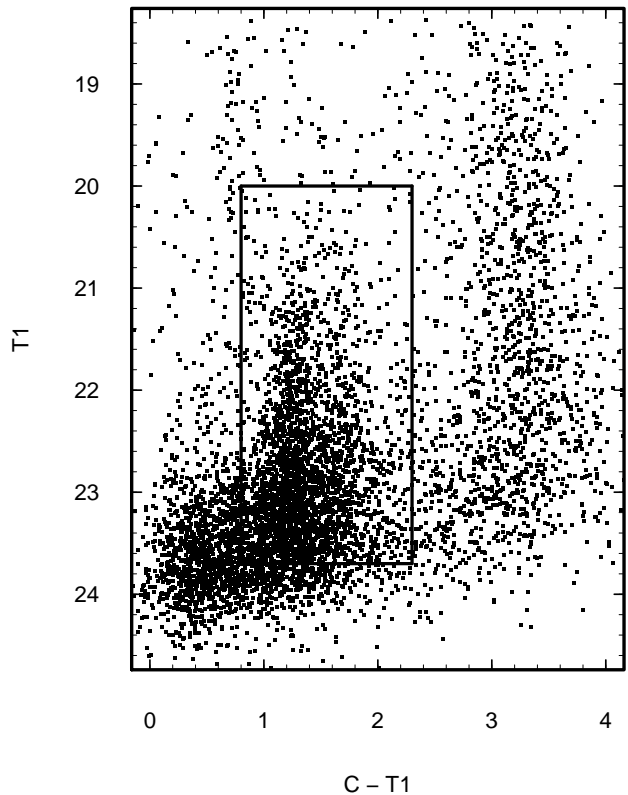}}
\caption{Colour-magnitude diagram for all point sources in the MOSAIC 
field. Globular cluster candidates show up within the box defined by 
the colour range $0.8 \leq (C-T_1) \leq 2.3$ and magnitude range 
$20 \leq T_1 \leq 23.7$.
}
\end{figure}

We select as GC candidates those point sources with colours
$0.8 \leq (C-T_1) \leq 2.3$ and magnitudes $20 \leq T_1 \leq 23.7$ 
where the faint $T_1$ limit is adopted according to the completeness 
results described in Section 2.3. All GC candidates identified on the 
MOSAIC field are shown in Fig.~4, where the GCSs around the three galaxies 
can clearly be seen; their positions, magnitudes and colours are 
listed in Table~2. We also notice in Fig.~4 a smooth background distribution of 
GC candidates with a projected density decreasing from East to West, 
representing the GCS of NGC\,1399, the central galaxy of the Fornax cluster. 
It is located at a
projected angular distance of 19 arcmin from NGC\,1387, towards the East.

\begin{figure}
\resizebox{\hsize}{!}{\includegraphics{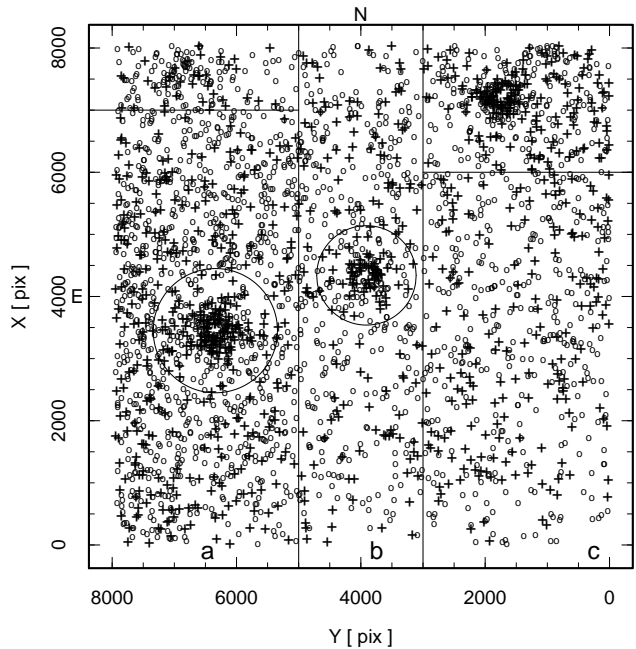}}
\caption{Projected distribution of GC candidates on the MOSAIC field:   
the colour $(C-T_1) = 1.45$ is adopted as limit between blues (open circles) 
and reds (crosses). 
The GCSs around the three target galaxies can be seen as well as 
the background distribution decreasing from East to West that corresponds to 
the NGC\,1399 GCS. Solid lines indicate the regions selected as background 
for each galaxy (labelled `a', `b' and `c') and large circles the regions 
excluded from them (see text). The MOSAIC scale is 0.27 arcsec\,pixel$^{-1}$. } 
\end{figure}

\begin{table}
\centering
\caption{Catalog of globular cluster candidates identified on the MOSAIC field. 
The full version of this table is available in the online article.} 
\begin{tabular}{cccc}
\hline
\noalign{\smallskip}
$\rmn {RA^\mathrm{a} (J2000)}$ & $\rmn {Dec.^\mathrm{a}(J2000)}$ & $T_1$ & $(C-T_1)$\\ 
\noalign{\smallskip}
\hline
$3^h 34^m 38\fs78 $ & $ -35\degr 19\arcmin 18\farcs6$ & $23.02 \pm 0.05$ & $2.22 \pm 0.09$\\
$3~~~34~~~39.55$   & $-35~~17~~27.4$ & $21.78 \pm 0.03$ & $2.11 \pm 0.04$\\
$3~~~34~~~40.64$   & $-35~~23~~01.7$ & $21.52 \pm 0.02$ & $1.37 \pm 0.02$\\
\hline
\end{tabular}
\begin{list}{}{}
\item[$^{\mathrm{a}}$] The estimated precision is 0.3 arcsec.
\end{list}
\end{table}

The selection of GC candidates should be corrected for  
the contamination by the background. In the study of the NGC\,1399 GCS 
we have used for this purpose a background field located as far as
3\fdg5 from the parent galaxy \citep{dir03a}. In the present case, 
where there is a rather 
strong contamination from NGC\,1399 itself (see Fig.~4) we 
must estimate the background contribution in regions closer to the target
galaxies, i.e. within the same MOSAIC image.
We should take into account that blue clusters from the GCS 
of NGC\,1399 extend up to 40 -- 45 arcmin~from the Fornax cluster 
centre (Bassino et al. 2005, in preparation) and that NGC\,1374, the target 
galaxy located farther away from it, has an angular separation of 41 arcmin. 
Thus, we have chosen as background 
for NGC\,1374 the region (labelled `c' in Fig.~4) located at  $y\,<$\,3000\,pix 
and $x\,<$\,6000\,pix  
(density $1.42 \pm 0.06$~objects\,arcmin$^{-2}$) where we  excluded 
the region with $y\,<$\,3000 and $x\,>$\,6000\,pix due to the presence of 
NGC\,1373, a small galaxy at 5 arcmin to the NW with respect to NGC\,1374 
(see Fig.~1), and whose probably own cluster candidates show up at 
$x$ = 8000 and $y$ = 1000 in Fig.~4. The background defined  
for NGC\,1379 is the region (labelled `b') located at 
3000\,$<\,y\,<$\,5000\,pix excluding a 
circle of radius 800\,pix (density $1.53 \pm 0.07$~objects\,arcmin$^{-2}$); 
for NGC\,1387 the region (labelled `a') located 
at $x\,<$\,7000 and $y\,>$\,5000\,pix  
excluding a circle of radius 1000\,pix (density $2.79 \pm 
0.09$~objects\,arcmin$^{-2}$) was selected as background; the region with  
$x\,>$\,7000 and $y\,>$\,5000\,pix 
was excluded due to the presence of another galaxy, NGC\,1380B. 

All these regions are shown in Fig.~4. It can be seen that the 
background density increases from West to East as we get closer 
to NGC\,1399. This effect is also present in the $V$ background counts 
obtained for the same galaxies by \citet[ see their fig.~1]{koh96}, although 
the influence of NGC\,1399 GCs is not mentioned as a background component.  

The reddening towards our three galaxies  taken from \citet{sch98},
is depicted in Table~1. The colour excess in the Washington photometric  
system $E(C-T_1)$ is estimated by the relation $E(C-T_1) = 1.97 E(B-V)$ 
\citep{har77}, and the absorption in $R$ with $A_R/A_V = 0.75$ \citep{rie85}. 
For the reddening of the three background regions defined in the 
previous paragraph, we adopt the values corresponding to the galaxies located 
in each of them, respectively. In the rest of the paper, all magnitudes and 
colours will be reddening free.

%-----------------------------------------------------------------------
\subsection{Completeness}

In order to estimate the completeness of the GC candidates for the $R$ 
and $C$ final images, we added 1\,000 artificial stars by means of the 
ADDSTAR task within DAOPHOT, which were generated with random $x,y$ 
positions scattered through the whole MOSAIC field. They were distributed 
within approximate the same the colour range used to select the GC 
candidates. This process was repeated 10 times, with 10 different 
seed numbers. The whole reduction process then was applied to 
each frame with added stars in exactly the same way as to  
the science images. The differences between input and output magnitudes 
and colours for the `added' stars did not show any systematic trend with 
the magnitude.

As a result, the completeness for stars in the colour range corresponding 
to the GC candidates is close to  90 per cent for a limiting magnitude $T_1=23$   
decreases rapidly to 50 per cent for a limiting magnitude $T_1=23.7$; 
this latter magnitude has been adopted as the faint magnitude limit 
of the GC candidates for further study in this paper. No 
difference was found in the completeness function for blue and red GCs, 
taking $(C-T_1)= 1.45$ as the colour limit between them. 

The analysis of the regions close to the galaxies' centres shows that 
inside $r$ = 100 pix. (27 arcsec) the detection of GC candidates is 
affected by the light of the galaxies and saturation so we will not 
consider those inner regions in our study. The areas corresponding to 
100 pix $< r <$ 200 pix (27 -- 54 arcsec) are still 
affected by the light of the galaxies and we expect there a completeness 
factor around 80 $\pm$ 5 per cent. The GC radial distributions will be 
corrected for this last effect and, in the following, we will refer to 
it as the innermost annulus or bin `geometric incompleteness'.

%=========================================================================
\section{The Globular Cluster Systems}

In what follows we will analyse the GCSs of the three galaxies.
On the basis of the projected distributions depicted in Fig.~4
and the radial density profiles (see below), the radial extensions of the 
GCSs will be taken as $r$ = 3.1 arcmin (700\,pix) for NGC\,1387, 
$r$ = 2.7 arcmin (600\,pix) for NGC\,1379, and $r$ = 2.2 arcmin (500\,pix) 
for NGC\,1374. We adopt an inner limit of at $r$ = 27 arcsec (100\,pix) for 
all GCSs (see Section 2.3).
The edge-on disc galaxy NGC\,1375 is seen in projection close to NGC\,1374 so 
its GCs may be contaminating the NGC\,1374 GCS; this is not causing a problem 
because the NGC\,1375 GCS does not seem to be very populous, as 
proved by inspection of the images.  NGC\,1373 seems
to have its own cluster system, as mentioned above, so we restrict 
the selection of GC candidates in NGC\,1374 to a radial distance of 500\,pix 
to exclude it. Table~3 gives the number of blue 
and red cluster candidates found in the regions defined for each galaxy. 
The colour range is  $0.8 \leq (C-T_1) \leq 2.3$  and the magnitude range
$20 \leq T_1 \leq 23.7$, 
 The selected colour limit between blue 
and red clusters is $(C-T_1) = 1.45$, as supported by the colour distributions 
discussed below (Fig.~6). 
We remind here that the adopted faint magnitude cut-off is $T_1 = 23.7$, the 
magnitude at which the completeness function is 50 per cent.
%----------------------------------------------------------------------
\subsection{Colour distributions}

Fig.~5 presents the colour-magnitude diagrams 
for point sources (selected as explained in Section 2.1) 
around the three galaxies within the 
radial limits stated above. GC candidates show up in them: the presence 
of two GC populations is clear in NGC\,1387 and NGC\,1374. The separation of 
blue and red GC candidates is not so clear in NGC\,1379. 
 
\begin{figure}
\resizebox{\hsize}{!}{\includegraphics{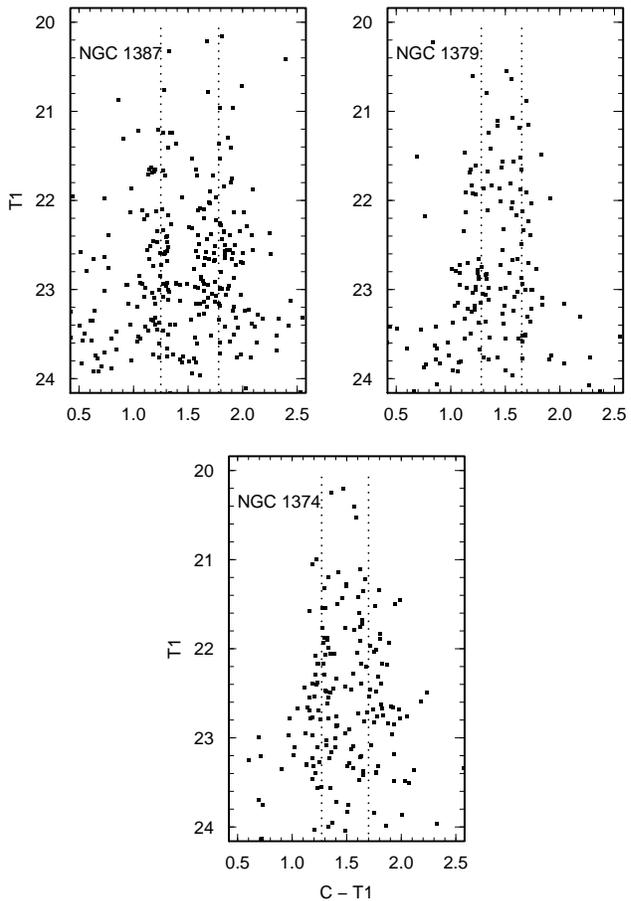}}
\caption{Colour-magnitude diagrams for point sources around the
galaxies indicated in each figure. Vertical lines indicate mean 
colours corresponding to the blue and red GC candidates, respectively, 
taken from Table~3.
}
\end{figure}

\begin{figure}
\resizebox{\hsize}{!}{\includegraphics{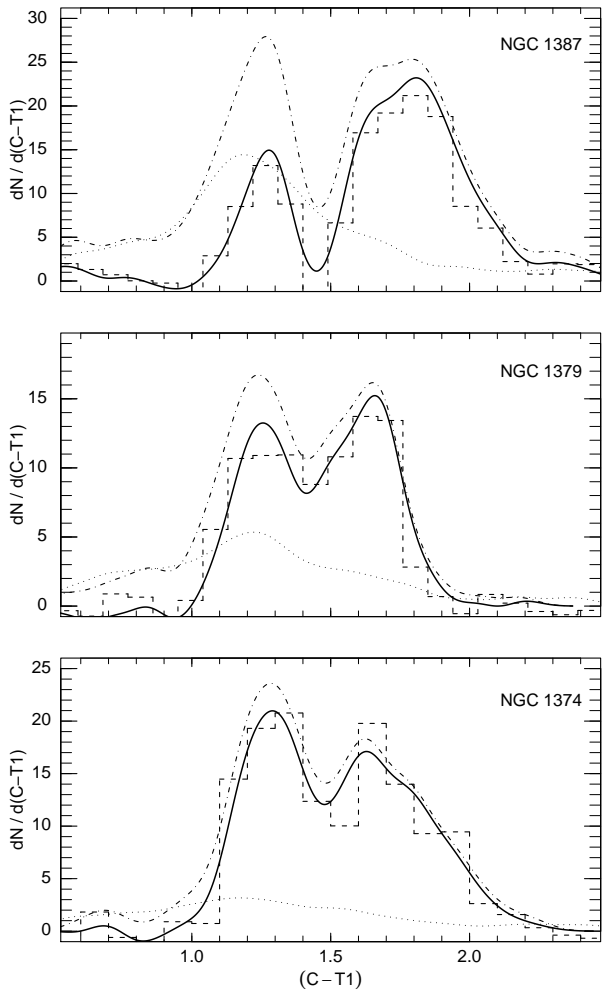}}
\caption{Colour distribution functions for the GC candidates around NGC\,1387 
(upper panel), NGC\,1379 (middle panel), and NGC\,1374 (bottom panel). 
Dash-dotted lines show the raw colour distributions and solid lines the 
background-corrected ones. The histograms of the data are 
plotted with dashed lines and the background colour distributions with 
dotted lines. Note that the displayed colour distributions and histograms span 
over a colour range slightly larger than the one adopted for the GC selection  
(i.e. $0.8 \leq (C-T_1) \leq 2.3$). 
}
\end{figure}

We can get a better picture from the colour distributions shown in 
Fig.~6. Displayed are the raw colour distributions, the backgrounds,  
and the background corrected distribution, all smoothed by  a
 Gaussian kernel with dispersion 0.09 mag. 
The histograms of the background-corrected data are also plotted.   
Not only the GCSs of  NGC\,1387 and 
NGC\,1374 appear clearly bimodal, but the one of NGC\,1379 too. It also 
can be seen in Fig.~6 that the contamination by the background decreases 
with increasing angular distance from NGC\,1399 (angular distances 
of NGC\,1387, NGC\,1379 and NGC\,1374 from NGC\,1399 are 19, 29 and 41 
arcmin, respectively) and that this effect is mostly due to a blue 
population. This is in agreement with the less concentrated 
radial distribution of blue clusters found at large distances from the 
parent galaxy (Bassino et al. 2005, in preparation) with respect to the 
radial distribution of red GCs.
In addition,  we note that the colour distribution of the background selected 
for NGC\,1399 by \citet[ their Fig.~6]{dir03a}, which is a field 
located 3\fdg5 northeast of the Fornax centre,  
looks very similar to the background chosen for 
NGC\,1374, the most distant galaxy from NGC\,1399.  
The backgrounds for NGC\,1379 and NGC\,1387 present larger blue populations, 
demonstrating the large extent of the NGC\,1399 system.

\begin{table*}
\centering
\caption{Number of identified GC candidates ($0.8 \leq (C-T_1) \leq 2.3$ 
and $20 \leq T_1 \leq 23.7$), and results of the Gaussian fits to their 
background-corrected colour distributions.}
\begin{tabular}{lcccccccc}
\hline
\noalign{\smallskip}
Galaxy & $\rmn N_{\rmn {blue}}^{~~\mathrm{a}}$  & $\rmn N_{\rmn {blue}}^{\rmn {corr}~~\mathrm{b}}$  & 
$(C-T_1)_{\rmn {blue}}$ & $\sigma_{\rmn {blue}}$ & $\rmn N_{\rmn {red}}^{~\mathrm{a}}$ & 
$\rmn N_{\rmn {red}}^{\rmn {corr}~~\mathrm{b}}$  &$(C-T_1)_{\rmn {red}}$ & $\sigma_{\rmn {red}}$\\
\noalign{\smallskip}
\hline
NGC 1374                & $75 \pm ~9$ & $61 \pm ~9$ & $1.27 \pm 0.02$ & 
$0.11 \pm 0.02$ & $~81 \pm ~9$ & $~73 \pm ~9$ & $1.70 \pm 0.03$ & $0.19 
\pm 0.03$\\
NGC 1379                & $64 \pm ~8$ & $40 \pm ~8$ & $1.28 \pm 0.02$ & 
$0.15 \pm 0.02$ & $~59 \pm ~8$ & $~49 \pm ~7$ & $1.65 \pm 0.01$ & $0.10 
\pm 0.01$\\
NGC 1387                & $94 \pm 10$ & $32 \pm 10$ & $1.25 \pm 0.01$ & 
$0.08 \pm 0.01$ & $123 \pm 11$ & $100 \pm 11$ & $1.78 \pm 0.01$ & $0.16 
\pm 0.01$ \\

\hline
\end{tabular}
\begin{list}{}{}
\item[$^{\mathrm{a}}$] Number of blue\,/\,red GC candidates without background
corrections and Poisson errors.
\item[$^{\mathrm{b}}$] Number of blue\,/\,red GC candidates 
background-corrected and Poisson errors including the background errors.
\end{list}
\end{table*}

Apparently, the better metallicity resolution provided by the Washington 
photometric system allows us to detect bimodalities that were not visible   
in other photometric systems. The GCSs of NGC\,1387, NGC\,1379 and 
NGC\,1374 did not show evidence for multiple populations in the $(V-I)$ 
photometry by \citet{kis97a}, neither did the $(B-I)$ 
photometry of NGC\,1379 presented by \citet{els98}, as already mentioned in the 
introduction. We are aware of the low metallicity resolution of the 
$(V-I)$ colour index \citep[see, for instance, fig.~7 in][]{dir03a} 
but one may wonder why the $(B-I)$ data, known to provide a 
better metallicity resolution than $(V-I)$ colours, do not show any 
evidence of bimodality for 
NGC\,1379 either. In order to analyse the colour distributions, the background 
corrections play a fundamental role. \citet{els98} used for their HST 
observations a background field lying 1\fdg4 to the south of the centre of 
the Fornax cluster;
it seems that this field probably lacks the strong contribution added  
to the background by the NGC\,1399 GCS. Also, in the same paper, there is no 
background field for the CTIO (ground-based) observations, so the authors use 
a corrected sample of the CFRS \citep[Canada--France Redshift Survey,][]{lil95} 
which might probably present the same problem. 

In addition to the individual GCS colour distributions of NGC\,1374, NGC\,1379, NGC\,1387 
and NGC\,1427, \citet{kis97a} presented,a $(V-I)$ composite histogram for all GCs with 
colour uncertainties less than 0.1 mag. Interestingly, this composite distribution 
seems to be 
bimodal according to a KMM test (the GCS of NGC\,1427 
has already been found to be clearly bimodal on the basis of the Washington 
photometry performed by \citet{for01}).

The GCSs of NGC\,1387 and NGC\,1374 are more populous than the one of 
NGC\,1379 within the quoted ranges of  
magnitude, colour, and radial distance (it should be noticed that vertical 
scales are different in Fig.~6). The number of identified red 
cluster candidates in NGC\,1387 outnumber the 
blue ones by more than a factor of three. They also reach redder colours as compared 
to the other GCSs. 

From Figs.~5 and 6 and from the data depicted in Table~3, the GCS 
of NGC\,1379 shows a similar number of blue GC candidates as red ones, 
although suggesting a larger number of red ones (40 blue/49 red). The same 
trend is present for the GC candidates around NGC\,1374, where the relative 
number of blue/red clusters results 61/73.

The fractions of blue/red cluster candidates for the three 
galaxies are 32 per cent for NGC\,1387,  82 per cent 
for NGC\,1379, and 84 per cent for NGC\,1374. 
Moreover, all three galaxies present more red clusters 
than blue ones, which is quite peculiar as compared to GCSs in 
general.  
One may suspect that either some kind of interaction might have been causing a loss 
of blue GCs from these galaxies, or that an enhanced cluster formation rate was responsible 
of an overabundance of red GCs in any of them, or both. We will come 
back to this point in the discussion. 
The results of the two-Gaussian fits applied to all colour distributions  
are also given in Table~3. The blue peaks agree within the errors and lie
in the range $(C-T_1) = 1.25 - 1.28$; it is possible to compare them 
with the blue peak colours determined for GCSs of other Fornax galaxies: 
$(C-T_1) = 1.32$ for GC candidates between $r$ = 1.8 and 4.5 arcmin in 
NGC\,1399 \citep{dir03a}, $(B-I) = 1.6$ for GCs with r $<$ 4 arcmin in 
NGC\,1404 \citep{for98} which corresponds to $(C-T_1) = 1.24$ applying the 
colour conversions proposed by \citet{forb01}, and $(C-T_1) = 1.35$ for 
candidates between $r$ = 0.5 and 5 arcmin in NGC\,1427 \citep{for01}.
They all agree among each other within 0.1 mag.    

On the other hand, the red peaks of the three target galaxies show larger 
differences among each other. 
Their range is $(C-T_1) = 1.65 - 1.78$, with a bluer peak
for NGC\,1379. The red peaks (within the same 
radial ranges and obtained by the same authors) have $(C-T_1) = 1.79$ for 
NGC\,1399, 
$(B-I) \approx 2.1$ which corresponds to $(C-T_1) = 1.76$ for NGC\,1404, 
and $(C-T_1) = 1.76$ for NGC\,1427. They agree with the NGC\,1387 value 
while peaks for NGC\,1374 and NGC\,1379 are bluer, more noticeable for 
NGC\,1379.

%--------------------------------------------------------------------- 
\subsection{Spatial distributions}

\subsubsection{Azimuthal distributions}

In order to analyse the radial density distributions of the GCSs, it should 
first be determined if they are spherical or if they show some 
ellipticity. Fig.~7 displays the azimuthal distributions of GCs for 
the three systems with respect to position angle (measured from 
North to East). 
No evidence for ellipticity can be seen in case of the NGC\,1387 system. 
The GCS in NGC\,1379 does not show any clear azimuthal dependence either 
within the uncertainties; as this galaxy is classified as E0, this is what one expects.
 
\begin{figure}
\resizebox{\hsize}{!}{\includegraphics{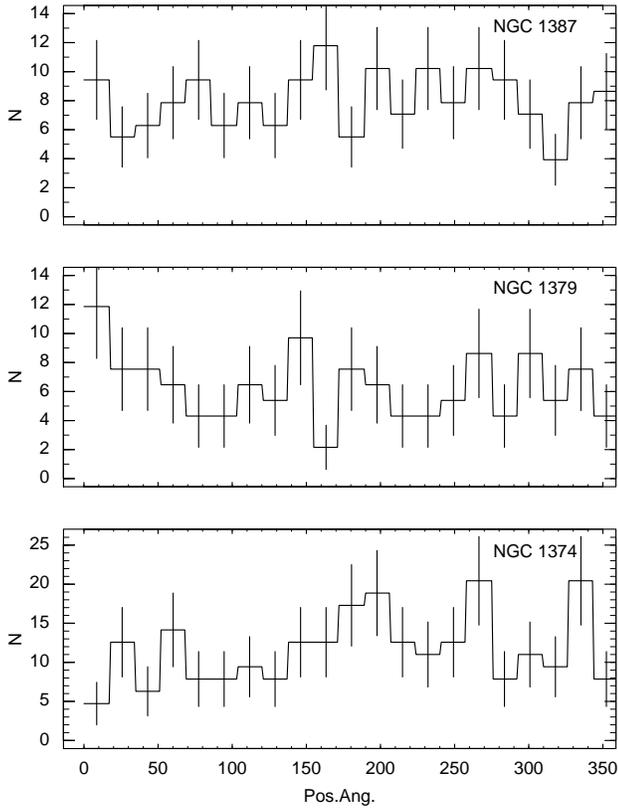}}
\caption{Azimuthal distribution of GC candidates in NGC\,1387 (upper panel),  
NGC\,1379 (middle panel) and NGC\,1374 (lower panel). Histograms bins span 
over $\approx 17\degr$ and Poisson errors are shown for them.
}
\end{figure}

The same holds true for NGC\,1374. It is classified as an E1 galaxy and has a 
ellipticity smaller than 0.1 as calculated with the ELLIPSE task 
within IRAF. So we expect again sphericity and all we can see in 
Fig.~7 seems to be an excess of GC candidates for 
position angles greater than $180\degr$ which is probably due to GCs from 
NGC\,1375. 

These results are consistent with those of \citet{kis97a}. 

%--------------------------------------------------------------------- 
\subsubsection{Radial distributions}

Radial number density profiles of the GCSs are presented in Fig.~8; 
they include the profiles uncorrected for  
background contamination as well as the background-corrected profiles. For the latter 
case, blue and red candidates are shown separately and are tabulated in 
Table~4. In all cases, the Poisson errors include the uncertainties  of the raw 
counts and of the background. 
The $ T_1 $ luminosity profiles of the corresponding galaxies, in a 
proper scaling, are also shown as well as the densities of background objects  
determined in Section 2.2.
We remind that the adopted cut-off magnitude is $T_1 = 23.7$. 

\begin{figure}
\resizebox{\hsize}{!}{\includegraphics{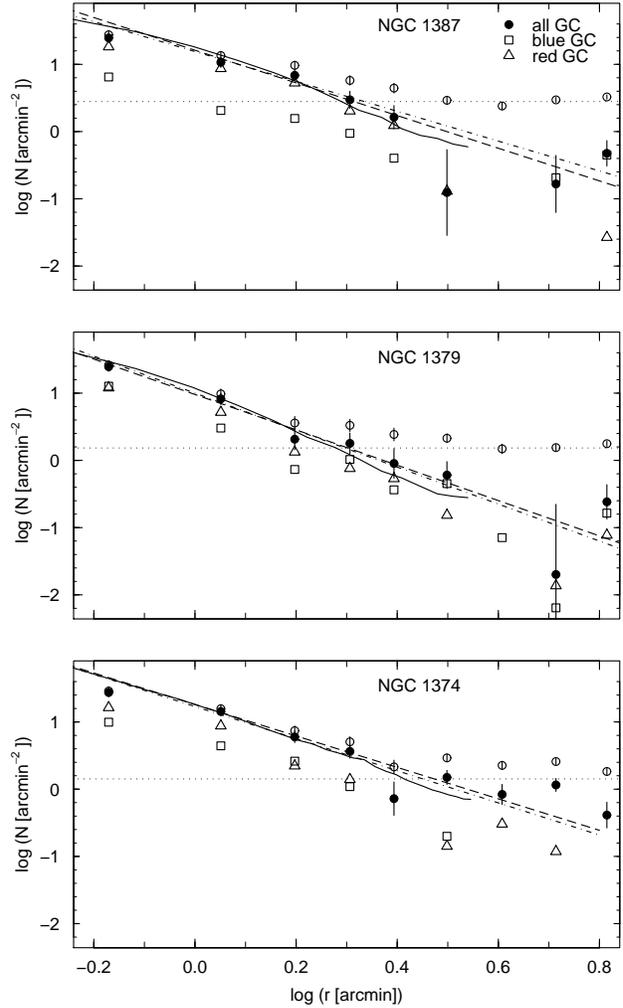}}
\caption{Radial density profiles for GC candidates in the field of NGC\,1387 
(upper panel), NGC\,1379 (middle panel), and NGC\,1374 (lower panel). Open 
circles show the radial distributions uncorrected for background 
contamination  and filled circles the 
background-corrected distributions, which are 
split in blue candidates (open squares) and red candidates (open triangles). 
Solid lines represent the $ T_1 $ brightness profiles of the galaxies 
multiplied by $-$0.4 and shifted by an arbitrary constant, dashed lines are 
the power-law fits to the GC radial density profiles, dot-dashed lines are 
the power-law fits to the galaxy brightness profiles, and dotted lines 
indicate the background densities. The innermost bins have been corrected by  
geometric incompleteness.
}
\end{figure}

\begin{table*}
\centering
\caption{Radial number densities of blue and red GC candidates 
(in arcmin$^{-2}$, and background-corrected) in the galaxies NGC\,1374, 
NGC\,1379 and NGC\,1387.}
\begin{tabular}{ccccccc}
\hline
\noalign{\smallskip}
r~[arcmin] & $\rmn N^{~\rmn {blue}}_{\rmn {NGC\,1374}}$ & $\rmn N^{~\rmn {red}}_{\rmn {NGC\,1374}}$ & $\rmn N^{~\rmn {blue}}_{\rmn {NGC\,1379}}$ & $\rmn N^{~\rmn {red}}_{\rmn {NGC\,1379}}$ & $\rmn N^{~\rmn {blue}}_{\rmn {NGC\,1387}}$ & $\rmn N^{~\rmn {red}}_{\rmn {NGC\,1387}}$\\
\noalign{\smallskip}
\hline
0.45 - 0.90 & $9.92\pm2.78$ & $16.34\pm3.34$ &  $12.69\pm3.00$ & $11.98\pm2.86$ & $6.49\pm2.36$ & $18.22\pm3.53$\\
0.90 - 1.35 & $4.41\pm1.41$ &  $8.73\pm1.72$ &  $3.02\pm1.14$ & $5.19\pm1.33$ & $2.06\pm1.14$ &  $8.66\pm1.72$\\
1.35 - 1.80 & $2.61\pm1.01$ &  $2.22\pm0.81$ &  $0.73\pm0.64$ & $1.33\pm0.64$ & $1.57\pm0.90$ &  $5.29\pm1.17$\\
1.80 - 2.25 & $1.09\pm0.72$ &  $1.40\pm0.61$ &  $1.03\pm0.61$ & $0.76\pm0.46$ & $0.94\pm0.72$ &  $2.03\pm0.70$\\
2.25 - 2.70 & $0.0 \pm0.50$ &  $0.00\pm0.25$ &  $0.36\pm0.46$ & $0.54\pm0.38$ & $0.40\pm0.59$ &  $1.23\pm0.54$\\
2.70 - 3.60 & $0.20\pm0.35$ &  $0.14\pm0.22$ &  $0.45\pm0.30$ & $0.15\pm0.19$ & $0.0 \pm0.35$ &  $0.13\pm0.23$\\
3.60 - 4.50 & $0.0 \pm0.26$ &  $0.30\pm0.22$ &  $0.07\pm0.23$ & $0.0 \pm0.13$ & $0.0 \pm0.28$ &  $0.0 \pm0.19$\\
4.50 - 5.85 & $0.0 \pm0.24$ &  $0.12\pm0.16$ &  $0.01\pm0.17$ & $0.01\pm0.11$ & $0.21\pm0.24$ &  $0.0 \pm0.14$\\
5.85 - 7.20 & $0.0 \pm0.18$ &  $0.0 \pm0.14$ &  $0.16\pm0.16$ & $0.08\pm0.11$ & $0.45\pm0.22$ &  $0.03\pm0.13$\\
\hline
\end{tabular}
\end{table*}

The innermost bins, which correspond to radial distances 100 $<$ r $<$ 
200 pix (0.45 -- 0.68 arcmin) from the galaxy centres, are affected by 
the light of the galaxies so we apply an 80 per cent completeness factor to 
correct for this `geometric' incompleteness.   

The density profile of NGC\,1387 clusters shows that the raw counts 
profile flattens at a level that agrees with the background density 
estimated above. The background-corrected density profile was fit 
by a power law of the form $N=a.r^{b}$ in the range 54 arcsec 
to 162 arcsec (2nd. to 5th. bins in Table~4), excluding the innermost 
bin and the outer regions where there are very few candidates left. 
The slope obtained 
for all GC candidates was $b = -2.4 \pm 0.3$, while for blue and red 
candidates separately they were $b = -2.0 \pm 0.5$ (blue clusters) and  
$b = -2.6 \pm 0.3$ (red clusters). The GC density profile in NGC\,1387 is clearly 
dominated by the red clusters, which show a steeper profile than the blue 
candidates, a result in common with other GCSs as, for example, the ones 
in the already mentioned Fornax early type galaxies NGC\,1399
\citep{dir03a}, NGC\,1404 \citep{for98}, and NGC\,1427 \citep{for01}. 
The three profiles (for all, red and blue candidates) are also fit 
reasonably well by a law $log N = c + d.r^{1/4}$~, within the same range 
used for the power law. 

The density profiles of the GC population in NGC\,1379 look similar to the 
ones for NGC\,1387, except that no dominant subpopulation is present in 
this case as red candidates are only slightly more abundant than blue ones. 
The raw counts profile levels out at the density value corresponding to the 
background objects, estimated near this galaxy. The background-corrected 
profile can be fit by a power law of slope  $b = -2.6 \pm 
0.5$ in the same radial range as in NGC\,1387 (2nd. to 5th. bins in Table~4); 
from the fits of blue and red candidates we obtain $b = -2.3 \pm 0.8$ 
(blues) and $b = -2.9 \pm 0.4$ (reds). The red clusters seem to be again more 
concentrated than the blue clusters, although the uncertainty  is rather 
large.
As in the case of NGC\,1387, an $r^{1/4}$ law provides a good fit for the 
density profiles of all candidates together as well as for blues and reds
separately.  

Finally, the GC density profiles for NGC\,1374 present a different behaviour, 
as depicted in Fig.~8. The raw counts profile does not flatten at the 
density value of the nearby estimated background objects due to the presence of 
GCs which probably belong to neighbouring galaxies; the candidates from 
NGC\,1375 may be the responsible of the excess at $r >$ 162 arcsec or 600 pix 
(see Table~4) while some candidates from NGC\,1373 may be also present at the 
two outer bins. 
To avoid this extra contamination we restrict the fit of the 
background-corrected counts by a power law to a smaller radial 
range 54 arcsec $< r <$ 135 arcsec (2nd. to 4th. bins in Table~4), from 
which we obtain slopes $b = -2.3 \pm 0.2$ (all candidates), $b = -2.3 \pm 0.6$ 
(blue clusters) and $b = -3.2 \pm 0.6$ (red clusters).    

The slope coefficients for the power law fits found by \citet{kis97a}, for 
all candidates, in NGC\,1387, NGC\,1379 and NGC\,1374 are: $-$2.2, $-$2.1 
and $-$1.8, respectively; they agree within the uncertainties with our values 
except for NGC\,1374 for which they obtain a shallower slope, probably due to 
the influence of the nearby contaminating GCSs. 

It is difficult to compare our results for NGC\,1379 with the HST based ones from 
\citet{els98} due to the different spatial coverage, limiting magnitude, 
and comparison field  (see Section 3.1).
Elson et al. argue that the GCS is lost in the background at 80 arcsec while, on 
the other side, they are able to analyse its behaviour even within 
10 arcsec; instead, our density profile suggests that the GC density reaches 
the background level at about 200 arcsec while we cannot study the GCs 
inside 27 arcsec. \citet{cap01} assign the difference between the surface 
brightness profile of NGC\,1379 and its GCS profile in the inner region 
($r <$ 50 arcsec) to a loss of GCs caused by dynamical friction and tidal 
interaction with the galactic nucleus; in our case the innermost bin in the 
radial density profile of NGC\,1379, corrected for geometric incompleteness,  
does not show any noticeable difference with the galaxy light profile. 

For the three galaxies under study, the radial density profiles  
GC candidates follow the $T_1$ light profiles of the respective 
galaxies (at least out to the radial distance at which they become undistinguishable  
from the background), as already pointed out by \citet{kis97a}.
The power law indices obtained for them are $b = -2.2 \pm 0.1$ for NGC\,1387, 
$b = -2.7 \pm 0.1$ for NGC\,1379, and $b = -2.4 \pm 0.1$ for NGC\,1374.

%-----------------------------------------------------------------------
\subsection{Luminosity functions}

The luminosity functions for the GC candidates (GCLFs) in NGC\,1387, 
NGC\,1379 and NGC\,1374 are presented in Fig.~9. As the images used 
in this study are not very deep, and the completeness sets the faint 
magnitude limit at $T_1 = 23.7$, we are not able to reach the turn-over 
magnitudes (TOMs) of the GCLFs with the required precision 
to derive distances. Instead, we opt to investigate whether these TOMs 
show the universal behaviour. 

\begin{figure}
\resizebox{\hsize}{!}{\includegraphics{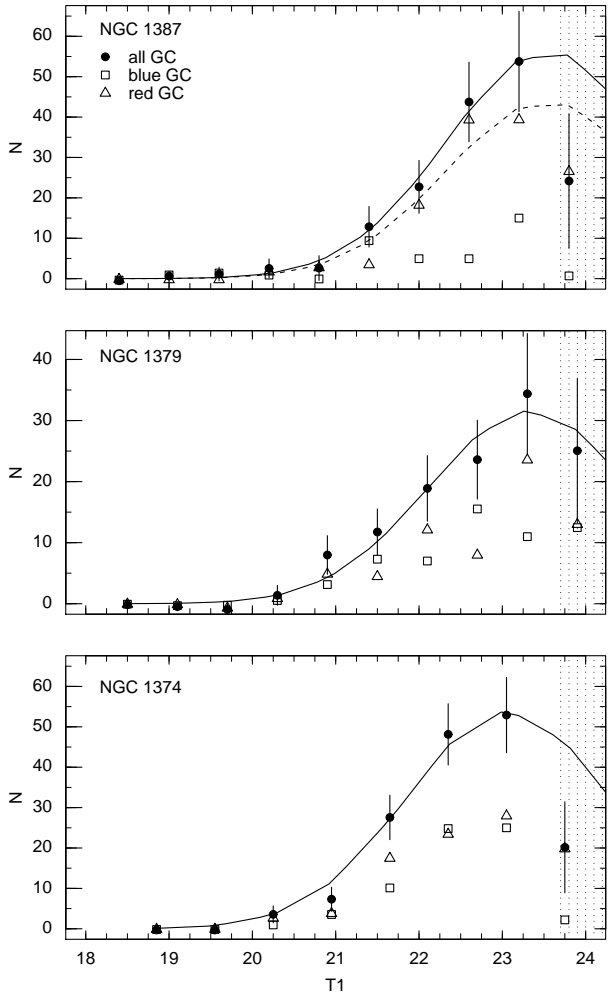}}
\caption{$T_1$ (dereddened) luminosity functions for the GC candidates 
in the field of NGC\,1387 (upper panel),  NGC\,1379 (middle panel), 
and NGC\,1374 (lower panel), for $r >$ 100pix (27 arcsec), and corrected 
for contamination by the background and photometric incompleteness. Solid 
lines show Gaussian fits to all candidates, and the dotted line the 
Gaussian fit only to the red candidates (open triangles); all fits were 
performed with a fixed dispersion $\sigma = 1.2$ and for magnitudes up to 
$T_1 = 23.7$, according to the 50 per cent completeness limit 
(in the shaded region the results are uncertain). Geometric incompleteness 
in the 
100 -- 200 pix annuli have not been corrected.  
}
\end{figure}

The GCLFs are based on the magnitude distributions corrected 
for photometric incompleteness and contamination by the background with 
bins of 0.6 mag 
for NGC\,1387 and NGC\,1379 and bins of 0.7 mag for NGC\,1374 for which we 
choose a slightly larger bin to diminish the errors as incompleteness seems 
to be affecting the fainter bins more in NGC\,1374 than in the other galaxies 
(see Fig.~9). As said above, an inner limit of $r$ = 27 arcsec is 
considered for all three GCSs and the geometric incompleteness in the 
100 -- 200 pix bins have not been corrected. 

As we are unable to construct the GCLFs beyond the TOMs, 
it is not possible to fit Gaussians leaving both mean value and 
dispersion as free parameters. We rather prefer to fix the dispersion 
at $\sigma = 1.2$, a value already obtained on the basis of 
Washington photometry by \citet{ost98} and \citet{dir03a} for NGC\,1399, 
and by \citet{for01} for NGC\,1427, and just estimate indicative TOMs 
by means of the Gaussian fits.   

The fit to all the NGC\,1387 GC candidates gives an indicative TOM  
at $T_1 = 23.54 \pm 0.06$, and for the red subpopulation at 
$T_1 = 23.53 \pm 0.18$, in agreement with the former value (more than 
75 per cent of the GCs belong to the red subpopulation in this galaxy). 
We can also estimate the expected TOM if we adopt the SBF 
distance modulus obtained by \citet{jen03} for this galaxy~ $(m-M) = 31.38 
\pm 0.26 $ and accept a universal TOM. The absolute TOM estimated by 
\citet{ric03} as a weighted average of the TOMs for the Milky Way and M31 
\citep[respectively]{har01,bar01} is $M_{V_0} = -7.46 \pm 0.18$.  
By means of the mean colour of the whole GC sample $(C - T_{1}) = 1.54 \pm 0.34$ 
and the colour conversions from \citet{forb01}, we can transform the 
$V$ band TOM into an $R$ band one $M_{R} = -8.00$. Thus, the expected apparent 
TOM is $R = 23.38$ (which corresponds to $T_1 = 23.40$ \citep{gei96a}); our 
observed value is 0.14 mag fainter.

In the case of NGC\,1379 and NGC\,1374, we perform fits 
to the whole population of GC candidates only, without separating between blue GCs
and red GCs. There are no dominant subpopulations and the number 
of only blue or red candidates would be too low. The indicative TOMs then 
result in $T_1 = 23.33 \pm 0.14$ for NGC\,1379, and $T_1 = 23.07 \pm 0.11$ 
for NGC\,1374. The SBF distances from \citet{jen03} are 
$(m-M) = 31.35 \pm 0.15 $ for  NGC\,1379 and $(m-M) = 31.32 \pm 0.13 $ for  
NGC\,1374. Following the same procedure, and by means of the mean 
colour of the GC populations: $(C - T_{1}) = 1.42 \pm 0.34$ for NGC\,1379 
and $(C - T_{1}) = 1.50 \pm 0.28$ for NGC\,1374, we estimate the expected 
absolute TOMs as $M_{R}= -7.97$ and $M_{R}= -7.99$, and the 
corresponding apparent TOMs as $T_1 = 23.40$ and $T_1 = 23.35$, for  NGC\,1379 
and NGC\,1374, respectively. For NGC\,1379 the observed and expected TOMs 
agree within the errors while for NGC\,1374 the observed TOM is about 
0.3 mag brighter. The observed TOMs for NGC\,1374 given by \citet{koh96} are  
$V = 23.52$ from a Gaussian fit and $V = 23.44$ from a $t_5$--function fit. 
The expected TOM according to the SBF distance is $ V = 23.84$, 
so in this case again the observed TOMs turn out to be 0.3 and 0.4 mag brighter, 
respectively, than the expected one.

%-----------------------------------------------------------------------
\subsection{Total populations and specific frequencies}

In order to calculate the specific frequencies we need the total GC populations
and the absolute V magnitudes of the galaxies.  
The total GC populations can be estimated from the luminosity functions, 
if we accept that they are fully Gaussian and the corresponding mean values 
and adopted dispersions are known. However, for all the GCSs under study, we 
lack information on the central 100 pix area, and also the surface density 
values for the first radial bin (100 pix -- 200 pix) are affected by  
geometric incompleteness that was not corrected in the GCLFs. We will then 
estimate, for each system, the population within the inner 100 pix assuming a 
constant 
surface density determined from the power law fit at $r$ = 100 pix; 
we justify this assumption for the three GCSs following \citet{els98} who 
found for NGC\,1379 that the surface density profile of the GCs flattens out 
inwards $r$ = 30 arcsec, which corresponds to 111 pix on our scale.  
In addition to this, a correction that should be taken into account is the 
difference between the observed radial density obtained for the 
first bin (uncorrected for the incompleteness) and the calculated 
radial density from the power law fit for the same bin.  
Thus, the inner 100 pix population and the correction for the 100 -- 200 pix 
range will be added, for each GCS, to the populations obtained from the GCLFs.

First, we calculate the raw number of GCs in the bright halves of the 
Gaussians (taking the TOM as upper magnitude limit), and correct 
them for contamination by the background and photometric incompleteness; 
these counts are then doubled to cover the whole Gaussian. The errors are 
calculated on the basis of the Poisson uncertainties of the raw and background 
counts, and the effect of the incompleteness. In this way, we obtain 
$274 \pm 18$ GCs for NGC\,1387, $160 \pm 11$ GCs for NGC\,1379, and 
$237 \pm 12$ GCs for NGC\,1374. Secondly, the fraction of the GC populations 
contained in the inner 100 pix corresponds to $72 \pm 15$ GCs for NGC\,1387, 
$50 \pm 15$ GCs for NGC\,1379, and $76 \pm 8$ GCs for NGC\,1374, where the 
errors take into account the errors of the fit. Finally, 
the corrections to be added due to the differences between observed and 
calculated densities in the first radial bins of each system, and their 
corresponding errors, are of $44 \pm 14$ GCs for NGC\,1387, $15 \pm 13$ GCs 
for NGC\,1379, and $47 \pm 9$ GCs for NGC\,1374. Thus, the total estimated 
populations result in $390 \pm 27$ GCs for NGC\,1387, $225 \pm 23$ 
for NGC\,1379, and $360 \pm 17$ GCs for NGC\,1374, where the uncertainties  
are estimated as a combination of the uncertainties  in the three steps just 
described; they are shown in Table~5.

\begin{table}
\centering
\caption{Total GC populations N$_{\rmn {GC}}$, integrated absolute 
magnitudes $M_V$ 
and specific frequencies S$_{\rmn N}$ for the three target galaxies.}
\begin{tabular}{lccc}
\hline
\noalign{\smallskip}
Galaxy & N$_{\rmn {GC}}$  & $M_{V}$ & S$_{\rmn N}$\\ 
\noalign{\smallskip}
\hline
NGC 1374                & $360 \pm 17$ & $-20.4 \pm 0.1$ & $2.4 \pm 0.5$\\
NGC 1379                & $225 \pm 23$ & $-20.6 \pm 0.2$ & $1.4 \pm 0.4$\\
NGC 1387                & $390 \pm 27$ & $-20.9 \pm 0.3$ & $1.8 \pm 0.7$\\
\hline
\end{tabular}
\end{table}

\citet{kis97a}  obtained $389 \pm 110$ GCs for NGC\,1387, $314 \pm 63$ GCs 
for NGC\,1379, $410 \pm 82$ GCs for NGC\,1374.  \citet{els98} 
quoted $436 \pm 30$ GCs for NGC\,1379. Our results for NGC\,1387 
and NGC\,1374 are in good agreement with \citet{kis97a}, particularly for 
NGC\,1387. In the case of NGC\,1379, our estimated GC population agrees 
barely within the errors with the one from Kissler--Patig et al., and it 
is about 50 per cent below the estimation from Elson et al. This large difference 
is likely a consequence of the different background: as the Elson et al. background 
is far away from NGC\,1399, it lacks the contribution of its CGS and this might 
turn into a overabundant GC population. 

The galaxies $T_{1}$ brightness profiles were integrated out to the respective 
radii adopted for each GCS and yield $T_{1} = 9.98 \pm 0.06$ for NGC\,1387 
($r <$ 3.1 arcmin), $T_{1} = 10.29 \pm 0.05$ for NGC\,1379 ($r <$ 2.7 arcmin), and 
$T_{1} = 10.39 \pm 0.04$ for NGC\,1374 ($r <$ 2.2 arcmin). The uncertainties of the 
integrated $T_{1}$ magnitudes arise basically  from the standard errors of the 
respective sky determinations.    
Our $T_{1}$ images are saturated at the three galaxy centres, over small 
regions with radii r $\approx$ 40 pix (about 0.2 arcmin). As we measure integrated 
magnitudes at larger limiting radii (r = 2.2, 2.7 and 3.1 arcmin), the proportion of 
light missed at the centres is negligible in comparison to the light integrated up to 
such limits. The $T_{1}$ integrated magnitudes were transformed to $R$ magnitudes 
through the zero-point difference ($R - T_1 \approx -0.02$) with the purpose of testing 
them against published $R$ multi-aperture photometry from \citet{pou88} and \citet{pou94}. 
In Fig.~10 we plot published and calculated integrated magnitudes for NGC\,1374, the galaxy 
with the smallest limiting radius, to show that the differences are evident 
at inner radii but tend to vanish for r $>$ 0.75 arcmin, that is, at a radius much 
shorter than the limiting one. We do not include the same plots for the other 
two galaxies as they look very similar. 

\begin{figure}
\resizebox{\hsize}{!}{\includegraphics{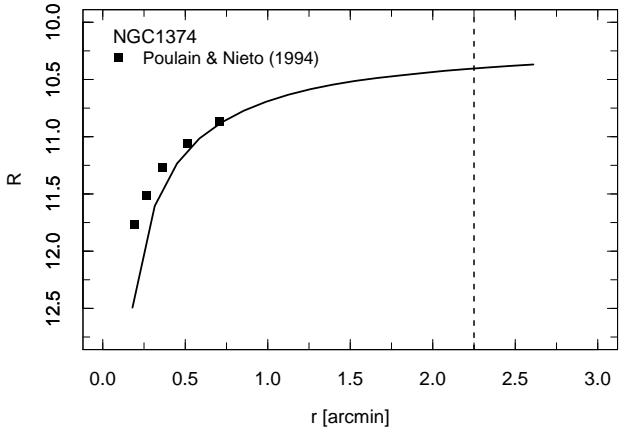}}
\caption{$R$ integrated magnitudes vs. radius for NGC\,1374. The solid line shows our 
results and solid squares plot the aperture photometry from \citet{pou94}. The vertical 
dashed line indicates the limiting radius where $R$ integrated magnitude has been 
measured for further calculations.}
\end{figure}

Adopting as representative colours for 
the galactic haloes of NGC\,1387, NGC\,1379 and NGC\,1374 the values obtained 
from the surface brightness profiles: $(C - T_{1}) = 1.61$, 1.49, and 1.45, 
respectively, and applying the colour conversions from \citet{forb01}, we 
calculate the corresponding $V$ magnitudes integrated up to those 
limiting radii as $V$ = 10.52, 10.80, and 10.89, respectively. 
By means of the adopted SBF distance moduli \citep{jen03}, we finally obtain the integrated  
absolute $V$ magnitude (up to the limiting radii) for NGC\,1387, NGC\,1379 and 
NGC\,1374: $M_{V} = -20.86 \pm$ 0.3, $ -20.55 \pm$ 0.2, and $-20.43 \pm$ 0.1, 
respectively,  whose uncertainties include the uncertainties  of 
the apparent integrated magnitudes and the distance moduli. In this way, 
the specific frequencies calculated with the GC populations and absolute 
$V$ magnitudes, both estimated up to the same radii, are $S_{N} =$ 1.8 $\pm$ 0.7 
for NGC\,1387, $S_{N} =$ 1.4 $\pm$ 0.4 for NGC\,1379, and $S_{N} =$ 2.4 $\pm$ 
0.5 for NGC\,1374; the uncertainties are calculated on the basis of the uncertainties 
of the absolute magnitudes and of the GC populations. Absolute $V$ 
magnitudes and specific frequencies are also given in Table~5.   
 
%========================================================================
\section{Discussion}

\subsection{Testing the tidal stripping scenario}  

The three galaxies under study present several characteristics that point to a 
particular interpretation. (i) \citet{har03} quote a `typical' specific 
frequency of $S_{N} \approx 4$ for elliptical galaxies but the values determined for 
NGC\,1379, and NGC\,1374 are distinctly smaller (NGC\,1387 is an S0 galaxy, where
a small $S_{N}$-value is not unusual)  (ii) The GCSs of these 
three galaxies show a rather small proportion of blue GCs with respect to the 
total GC population: 24 per cent for NGC\,1387, 45 per cent for NGC\,1379, and 
43 per cent for NGC\,1374. (iii) The radial distributions of blue clusters  
are less concentrated than the red ones with respect to the galaxy light
which implies a higher probability for the blues to get lost during tidal 
stripping processes. (iv) The galaxies are all located close to NGC\,1399, a giant  
elliptical galaxy with a very rich GCS. These properties make it likely  that these 
three galaxies have experienced tidal stripping of their blue GCs by NGC\,1399. 
\citet{kis99} already suggested that the rich cluster system of NGC\,1399 have partly 
been formed by tidal stripping of GCs from neighbouring galaxies. 

\citet{bek03} perform numerical simulations on the tidal stripping and accretion of GCs 
for the case NGC\,1404 and NGC\,1399 in Fornax. They propose to test the influence of  
tidal stripping in low $S_{N}$ ellipticals that might be involved in such process,  
looking for 'tidal streams' of intracluster GCs along the orbit of their former parent 
galaxy, and checking if there is some correlation between the distance of an elliptical 
galaxy from the center of the cluster and $S_{N}$. With regard to the first test, the GCs  
around NGC\,1399 present asymmetries in their azimuthal projected distribution and a 'bridge' 
of blue GCs in the direction to NGC\,1387 appear quite distinctly, pointing to an overdensity 
of GCs in the particular direction. We cannot assure that these GCs have been in fact lost 
by NGC\,1387 as its orbit around NGC\,1399 is not known, but being the blue globulars less 
bound to NGC\,1387 than the reds, due to their larger mean distance from the host galaxy,  
they have a higher probability of being lost. 

The other test proposed by \citet{bek03} is to search for a correlation between the $S_{N}$ 
of the elliptical galaxies and their projected distances to NGC\,1399. Fig.~11 shows 
such relation for the NGC\,1374 and NGC\,1379 as well as for NGC\,1404 and NGC\,1427. 
Other Fornax galaxies close to NGC\,1399 like NGC\,1380 and NGC\,1387 
are not included as they are lenticulars. There seems to be a trend in the sense that 
galaxies with larger projected distances from the cluster center tend to have higher 
specific frequencies, but there are only four points in the graph. Previous results 
related to this test have been presented by \citet{for97}. 
 
\begin{figure}
\resizebox{\hsize}{!}{\includegraphics{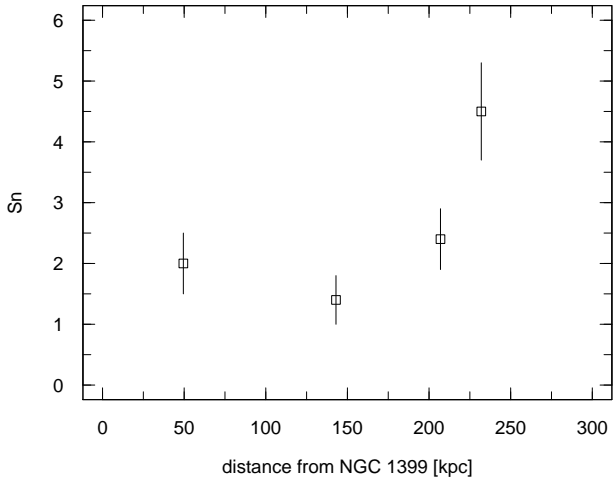}}
\caption{Specific frequency vs. projected distance from NGC\,1399. 
From left to right, the data are from NGC\,1404, NGC\,1379, NGC\,1374 
and NGC\,1427. The $S_{N}$ were taken from \citet{for98}, this paper and \citet{for01}, 
respectively.}
\end{figure} 

Further evidence for this may be expected from kinematical studies of the NGC\,1399 
system at radii larger than presently available velocities \citep{ric04}.

\subsection{The peculiar colour distribution of the GCS in NGC\,1387}  

The colour distribution of the NGC\,1387 GCS (see Fig.~6) is peculiar due to the 
large number of red clusters and also because they extend to redder colours 
than in most GCSs. One possibility to explain such distribution is the 
presence of background galaxies that might have been erroneously classified 
as red GC candidates. No contamination of this type is expected 
from the results of the background adopted (region `a' in Fig. 4): it can be 
seen in Fig. 6 that the contribution of objects redder than $(C - T_{1}) = 1.45$ 
is very low and does not affect the shape of the red GCs colour distribution. 
Even if such unresolved galaxies were present in the same area as the GCS, 
it is shown in Paper I that they would have contaminated the GCs colour range 
in $V-I$ but not in $(C - T_{1})$ as they appear at $(C - T_{1}) < 0.8$ 
(see Paper I, their fig. 7). Objects with $(C - T_{1}) < 0.8$ and $T_{1} > 23$ 
do not show any concentration towards the target galaxies, like the GC candidates, 
but they present an homogeneous projected distribution over the whole field 
considered in the present paper. Thus, we expect the contamination of the red 
candidates by background galaxies to be low.  

The lack of blue GCs may be explained by tidal stripping processes, 
as mentioned above, but NGC\,1387 is an S0 galaxy and alternative explanations 
may be valid as well.  
The GCS of the S0 galaxy NGC\,1380 in the Fornax cluster has been studied by 
\citet{kis97b} and it presents several properties similar to the ones we describe 
in the NGC\,1387 system. The colour distribution of NGC\,1380 GCs is clearly bimodal 
and their $(B-V)$ and $(B-R)$ colour histograms (their figs.~3 and 4) show a 
large number of red GCs that extend to redder colours than the metal-rich 
population in the Milky Way, like in NGC\,1387. The fraction of blue clusters 
over the total population is small (32 per cent) as well as the specific frequency 
obtained by Kissler--Patig et al. $S_{N} =$ 1.5 $\pm$ 0.5. These similarities 
between the GCSs of NGC\,1380 and NGC\,1387 are also evidence that the peculiar 
GCS colour distribution in NGC\,1387 is hardly due to background galaxies. 

On the other hand, the blue population of NGC\,1380 shows a spherical distribution 
around the galaxy while the red one follows the stellar light in ellipticity and 
position angle so they can be associated with the galactic bulge and disc. When 
we study separately the radial distributions of blue and red GCs in NGC\,1387, the 
blues show no sign of ellipticity and the reds only a marginal evidence, too weak 
to be significant. In this way, we cannot distinguish between a bulge and disk population.
The formation of S0 galaxies is still a matter of discussion. As a class, they are probably 
not simply gas-stripped spirals \citep{bur05}. However, interaction 
between galaxies is likely to play a role in their formation history 
\citep[e.g.][]{bic03,shi04, fal04}. 
This interaction could have triggered a starburst in NGC\,1387, causing the abundant 
formation of metal-rich clusters. Since NGC\,1387 is seen almost face-on, a cluster
disk population is not distinguishable from a halo population by its projected spatial 
distribution.
A radial velocity study could reveal whether the metal-rich clusters in NGC\,1387 show a disk
kinematical behaviour.  

%-----------------------------------------------------------------------
\subsection{The GCS colour distributions compared to the luminosities of the 
parent galaxies}
  
In order to analyse the trend of the $(C-T_1)$ colours of the blue and red peaks 
of GCSs with luminosity of the corresponding parent galaxies we present in 
Table~6 the data obtained from previous Washington photometric studies and 
from this paper. The sources are detailed in the table.

\begin{table}
\centering
\caption{Total visual magnitudes and $(C-T_1)$ colours of the blue and red GC peaks for 
galaxies with Washington photometry.}
\begin{tabular}{@{}l@{~~}c@{~~}c@{~~}c@{~~~}c@{~~~}c@{}}
\hline
\noalign{\smallskip}
Galaxy & $m-M^{\mathrm{a}}$ & ${M_{V}^{\rmn T}}^{\mathrm{b}}$ & $(C-T_1)_{\rmn {blue~pk}}$
 & $(C-T_1)_{\rmn {red~pk}}$ & Sour.$^{\mathrm{c}}$ \\ 
\noalign{\smallskip}
\hline
NGC\,1374 & 31.32 & -20.22  &  $1.27 \pm 0.02$ & $ 1.70 \pm 0.03$ & 1 \\ 
NGC\,1379 & 31.35 & -20.36  &  $1.28 \pm 0.02$ & $ 1.65 \pm 0.01$ & 1 \\
NGC\,1387 & 31.38 & -20.66  &  $1.25 \pm 0.01$ & $ 1.78 \pm 0.01$ & 1 \\
NGC\,1399 & 31.02 & -21.53  &  $1.32 \pm 0.05$ & $ 1.79 \pm 0.03$ & 2 \\
NGC\,1427 & 31.02 & -20.11  &  $1.35 \pm 0.07$ & $ 1.76 \pm 0.07$ & 3 \\
NGC\,3258 & 32.53 & -21.23  &  $1.36 \pm 0.04$ & $ 1.70 \pm 0.02$ & 4 \\
NGC\,3268 & 32.71 & -21.41  &  $1.31 \pm 0.02$ & $ 1.66 \pm 0.03$ & 4 \\
NGC\,3923 & 31.80 & -22.11  &  $1.47 \pm 0.025$ & $1.87 \pm 0.025$ & 5 \\
NGC\,4472 & 31.20 & -22.82  &  $1.31 \pm 0.04$ & $ 1.81 \pm 0.04$ & 6 \\
NGC\,4636 & 31.24 & -21.73  &  $1.28 \pm 0.02$ & $ 1.77 \pm 0.02$ & 7 \\
\hline
\end{tabular}
\begin{list}{}{}
\item[$^{\mathrm{a}}$] Distance moduli from GC luminosity functions 
(when available) or SBF distances \citep{jen99,ton01,jen03}.
\item[$^{\mathrm{b}}$] $M_{V}^{\rmn T}$ calculated with $V$ total magnitudes 
from RC3 catalog and the distance moduli quoted in the second column.
\item[$^{\mathrm{c}}$] $(C-T_1)$ peak colours sources. 1: this paper, 
2: \citet{dir03a}, 3: \citet{for01}, 4: \citet{dir03b}, 5: \citet{zep95}, 
6: \citet{gei96b}, 7: \citet{dir05}
\end{list}
\end{table}

Blue and red peak colours versus visual absolute magnitudes are plotted in Fig.~12. 
By means of  linear fits performed to the data we obtain a slope of $-0.02 \pm$ 0.02 
for the blue peaks and $-0.04 \pm$ 0.02 for the red ones.
The general appearance resembles that found in $(V-I)$ \citep{lar01,kun01a,kun01b}: 
the colour of the red peak depends slightly on the luminosity of the host galaxy,
while the blue peak shows a much shallower dependence, reasonably consistent with
being constant. The scatter is of the order which can be expected for a 
inhomogeneous sample and small number statistics cannot be ignored either. 
The only previous Washington photometric study that has not been included in the 
sample of GCSs depicted in Table~6 is NGC\,3311 as it seems to pose a problem. 
\citet{sec95} claimed that NGC\,3311 lacks the population of metal-poor clusters. 
However, \citet{bro00} found the $(V-I)$ colour distribution quite normal. Since the 
photometric quality of the Washington photometry was doubtful (Geisler, priv. communication) 
we excluded this GCS of the selected sample.

\begin{figure}
\resizebox{\hsize}{!}{\includegraphics{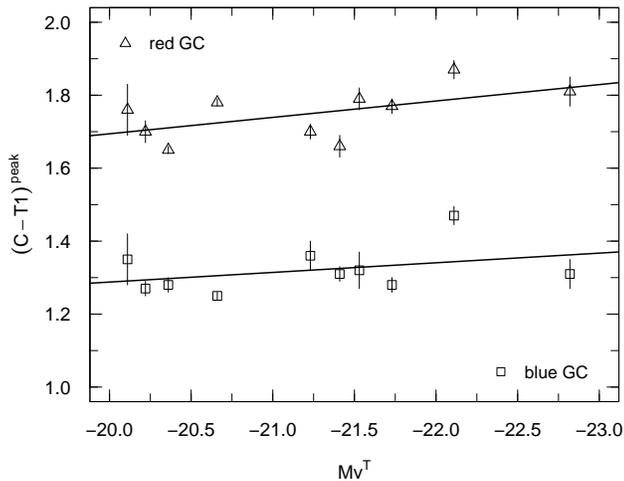}}
\caption{Blue and red $(C-T_1)$ peak colours of GC colour distributions 
in galaxies with Washington photometry vs. luminosity of the parent 
galaxies. The solid lines show the linear fits to the red and blue 
GC peaks, respectively.}
\end{figure}

That the red peaks have a slight dependence on the host galaxy's luminosity seems to be 
plausible: brighter galaxies possess more metal-rich clusters. The well-known relation 
between brightness and metallicity for early-type galaxies should be somehow reflected 
in the metallicities of the metal-rich clusters.   
The constancy of the blue peak has been a matter of discussion. \citet{lar01} find that
the blue peak becomes bluer with decreasing galaxy luminosity (by an amount of 0.016 mag in 
$(V-I)$ per B-magnitude of the host galaxy), while \citet{kun01a,kun01b} do not see 
this effect. 
Given the superior metallicity resolution of the Washington system over $(V-I)$, we
would expect to uncover a relation if there was any. On the other hand, a constant blue 
peak does not necessarily mean that all metal-poor GCs have the same metallicity 
distribution. As mentioned in \citet{dir03a}, a non-linear colour-metallicity relation in 
combination with a certain scatter around the mean relation (by second parameters or 
photometric errors) can produce a quite universal 
blue peak, if a given metallicity distribution is rebinned in colour. 
The colour-metallicity relation
for the Washington system is non-linear \citep{har02}. Due to its large scatter, the $(V-I)$  
colour-metallicity relation is difficult to
calibrate in the metal-poor regime of Galactic globular clusters, 
but it is reasonable to assume that it also becomes non-linear.  

This does not necessarily contradict the results of \citet{str04}, who found a  
significant relation between the {\it mean} blue colours in $(V-I)$ and galaxy luminosity. 
They  include very faint dwarf galaxies whose cluster systems contain only a few 
clusters and where the quotation of a `peak' is meaningless. 

%-----------------------------------------------------------------------
\section{Conclusions}

The colour distributions of the GCSs in the low-luminosity galaxies NGC\,1387, 
NGC\,1379, and NGC\,1374 show a distinctly bimodal appearance; 
these bimodalities have not been 
detected in previous work. NGC\,1387 even shows a colour bimodality more 
notably than known from any other GCS. That bimodality is not an exclusive 
property of the GCSs of giant elliptical galaxies, has already been shown 
in another Fornax low-luminosity galaxy: NGC\,1427 \citep{for01}.
Blue and red GC candidates also show different spatial 
distributions, the red GCs being more centrally concentrated than the blue ones. 
The appearance of bimodality in our Washington colours probably is a consequence of
the superior metallicity resolution of the Washington system with respect to other
broad band colours. In fact, {\it all} GCSs observed so far in the Washington
system revealed bimodal distributions, while in $(V-I)$, perhaps due to the inferior
metallicity resolution, a certain fraction remained unimodal or unclear 
\citep[e.g. ][]{lar01,kun01a,kun01b}. One may suspect that a bimodal colour distribution
is a property of {\it all} GCSs, as long as there is a statistically significant 
to use this concept. It is therefore desirable to investigate a larger sample of 
low-luminosity galaxies in the Washington system. 

With the aid of the luminosity functions we estimate, for these three low-luminosity 
galaxies, GC populations between 200 and 400 
GCs and specific frequencies in the range $S_{N}$ = 1.4 -- 2.4 , i.e. smaller than 
the typical values for elliptical galaxies \citep{har03}, though we remind here 
that NGC\,1387 is a S0 galaxy, not an elliptical. Though it is likely that 
the GCSs under study owe the characteristics of their GC colour distributions and 
low specific frequencies to some interaction processes, as discussed in the previous 
section, it is worth to analyze them, in the following, under the hypothesis that 
they are just regular systems whose properties have not been affected by, for instance, 
tidal stripping.  

The specific frequencies provide us with valuable information related to the 
efficiency of the galaxies at forming GCs as well as the role the environment 
might have played during the evolution. For instance, \citet{forb01b} found that 
specific frequencies of red GCs 
estimated with respect to (only) the 'bulge' luminosity of spiral galaxies $S_{N}$ = 
0.5 -- 1.5 are similar to the ones of red GCs in field ellipticals estimated 
in the usual way $S_{N}$ = 1 -- 3 \citep[see references in ][]{forb01b}. It is 
interesting to note that the specific frequencies we obtain for the red candidates 
around the ellipticals NGC\,1374 ($S_{N}^{red} = 1.4 \pm 0.3$) and NGC\,1379 
($S_{N}^{red} = 0.7 \pm 0.3$) are also close to these $S_{N}$ ranges      
which can be considered as an evidence in favour of the Forbes et al.  
claim that metal-rich GCs in spirals and ellipticals may have the same origin: 
they are formed with the bulge stars, assuming that ellipticals are bulge-dominated 
galaxies. The numerical simulations performed by \citet{bea02} on the formation 
of GCs within a hierarchical Universe also provide theoretical support to this 
statement.  

Let us turn now to the predictions particularly related to low-luminosity galaxies.  
\citet{for97} discussed different models for the origin of GCs in giant 
elliptical galaxies and suggested, as the most probable scenario, the formation 
of GCs 'in situ' in two star formation episodes. One of the predictions that 
results from the Forbes et al. analysis and can be tested through observations, 
is that low-luminosity, low $S_{N}$ galaxies should have more metal-rich than 
metal-poor GCs. This trend is not followed, for example, by NGC\,1427 
\citep[$M_V = -20.05$,][]{for01} with a blue/red ratio of GCs in the range 
3.9 -- 4.4, or by NGC\,3379 in the Leo I Group \citep[$M_V = -20.9$,][]{rho04} 
with a 70 \% of blue GCs with respect to the total globular population. Instead, 
the three low-luminosity galaxies under study present blue/red ratio in the range 
0.3 -- 0.8, in agreement with the Forbes at al. predictions. Anyway, we should 
keep in mind that the low number of blues in these GCSs may be originated in another 
processes, like tidal stripping, and that one of the galaxies is of S0 type. Thus, our 
blue/red GC ratios do not provide much evidence in favour of the proposed model.

According to our results, the `classic' interpretation of bimodality as a 
consequence of merger events which would add a metal-rich GC population to 
a pre-existing metal-poor GC populations, is doubtful in the case of 
low-luminosity ellipticals. However, the hierarchical scenario 
proposed through simulations by \citet{bea02}, with blue GCs 
formed in protogalactic cold-gas fragments and red GCs formed in 
subsequent gas-rich mergers, leads to bimodal colour distributions in 
galaxies of all luminosities, even low-luminosity ones. 
According to Beasley et al. predictions, the metallicity distributions 
of blue GCs are very similar in low and in more luminous ellipticals. The red 
GCs in low-luminosity ellipticals, as these galaxies suffer fewer mergers 
followed by star formation than the luminous ellipticals, present a larger 
scatter in their mean colours. Both predictions are verified here by the GCSs 
of the ellipticals NGC\,1374 and NGC\,1379.

After checking the properties of the GCSs around NGC\,1374, NGC\,1379 and 
NGC\,1387 against previous observational and theoretical results, we suggest 
that a plausible scenario of formation and evolution of GCSs which 
shows reasonably agreement with our results would be a combination of 
formation in a hierarchical Universe \citep{bea02} and further tidal stripping,  
or other sort of interaction processes \citep[][ for instance]{for97}, 
originated in their proximity to the giant elliptical NGC\,1399. 
We reiterate that kinematical studies of these GCSs would help to obtain 
more evidence on the proposed statements.
  
%-----------------------------------------------------------------------
\section*{Acknowledgments}

This work was funded with grants from Consejo Nacional de Investigaciones 
Cient\'{\i}ficas y T\'ecnicas de la Rep\'ublica Argentina, Agencia Nacional 
de Promoci\'on Cient\'{\i}fica Tecnol\'ogica and Universidad Nacional de La 
Plata (Argentina), and from the Chilean {\sl Centro de Astrof\'\i sica} 
FONDAP No. 15010003. LPB is grateful to the Astronomy Group at the Universidad de 
Concepci\'on for the financial support and warm hospitality.

\label{lastpage}
\end{document}